\documentstyle[aps,preprint,eqsecnum]{revtex}
\begin{document}
\draft
\tighten
%@twocolumnfalse\endcsname
\title{A point mass in an isotropic universe: III. The region $R\leq 2m$}
\author{Brien C. Nolan\footnote{e-mail: nolanb@ccmail.dcu.ie}\\
School of Mathematical Sciences,\\
Dublin City University,\\ Glasnevin, Dublin 9,\\
Ireland.}
%\date{\today}
\maketitle
%\newpage
\begin{abstract}
McVittie's solution of Einstein's field equations, representing a
point mass embedded into an isotropic universe, possesses a scalar
curvature singularity at proper radius $R=2m$. The singularity is
space-like and precedes, in the expanding case, all other events
in the space-time. It is shown here that this singularity is
gravitationally weak, and the possible structure of the region
$R\leq 2m$ is investigated. A characterization of this solution
which does not involve asymptotics is given.
\newline
\end{abstract}

\section{Introduction}
McVittie's solution of Einstein's field equations \cite{mcv} arises uniquely under the following hypotheses \cite{mcv1}:
\newcounter{conditions}
\begin{list}
{(\roman{conditions})}{\usecounter{conditions}}
\item Space-time $(M,g)$ is spherically symmetric and is filled with a shear-free perfect fluid.
\item $g\sim g_0$ as $R\rightarrow\infty$, where $R$ is the proper radius function of $(M,g)$ and $g_0$ is the line element of a Robertson-Walker (RW) universe.
\item The energy density of $(M,g)$ is spatially homogeneous;  $\rho=\rho(t)$.
\end{list}

In two previous papers, we studied various properties of space-times described by (i)-(iii). In \cite{mcv1} (paper I), we proved the uniqueness result referred to above and discussed basic properties of the resulting space-times. (The plural here refers to the two different cases which arise according as the RW background has flat $(k=0)$ or negatively curved $(k=-1)$ spatial sections. Since (ii) above refers to infinite spatial distances, the case $k=+1$ was not dealt with. That gap will be filled here.) In \cite{mcv2} (paper II),
we studied the global structure of McVittie's $k=0$ solution \cite{mcv}, which is the unique space-time arising in the case $k=0$.
The line element of this space-time may be written as
\begin{eqnarray}
ds^2 =-\left( \frac{1-m/2u}{1+m/2u}\right)^2dt^2+
e^{\beta(t)}(1+m/2u)^4(dr^2+r^2d\omega^2),\label{lel1}
\end{eqnarray}
where $u=re^{\beta/2}$, $m$ is constant and $d\omega^2$ is the line element of the unit 2-sphere. Notice that the case $m=0$ gives a $k=0$ RW universe, while $\beta=$constant gives the isotropic form of the Schwarzschild solution. The coordinate ranges must be carefully considered, for the radius function
\begin{eqnarray}
R(u)=u(1+m/2u)^2 \label{req}
\end{eqnarray}
is double valued on the interval $u\in (0,\infty)$.
In fact we showed in paper II that $u\in(0,m/2]$
corresponds to $R\in[2m,\infty)$, yielding a solution
which represents a point mass embedded in a $k=0$ RW universe with
scale factor $S^2(t)=e^{-\beta(t)}$, while $u\in[m/2,\infty)$
corresponds to the same with scale factor $S^2(t)=e^{\beta(t)}$.
We deal with the latter case, and using the proper radius $R$ as a coordinate, the line element (\ref{lel1}) may be written
\begin{eqnarray}
ds^2=-(1-\frac{2m}{R}-\frac{1}{4}\beta_t^2R^2)dt^2-\beta_tR(1-\frac{2m}{R})^{-1/2}dRdt+(1-\frac{2m}{R})^{-1}dR^2+R^2d\omega^2.\label{lel}
\end{eqnarray}
The subscript indicates partial differentiation with respect to $t$. We refer to the space-time with this line element as $McV_4$.

Notice then that the line element is singular at $R=2m$. This is not a coordinate effect, as the form of the energy density and pressure (calculated via Einstein's field equations) show:
\begin{eqnarray}
8\pi\rho=\frac{3}{4}\beta_t^2,\qquad
8\pi p=-\frac{3}{4}\beta_t^2-\beta_{tt}(1-\frac{2m}{R})^{-1/2}.
\end{eqnarray}
Thus there is a scalar curvature singularity at $R=2m$. The expansion of the fluid flow lines is homogeneous, and equal to that of the RW background;
\[\theta(t)=\frac{3}{2}\beta_t(t).\]
Thus we may refer to a given $McV_4$ as expanding or contracting according as
$\theta$ is positive or negative. In paper II, we showed that in very general
circumstances, namely those which lead to a big bang (big crunch) for the
expanding (contracting) RW background, the surface $R=2m$ acts as a
space-like barrier which precedes (succeeds) all other events in the
space-time. More precisely, we showed that all radial null geodesics
originate (terminate) at $R=2m$. Thus this singular surface shuts off
the region $R<2m$ from the rest of the space-time.

In fact, the coordinates of neither (\ref{lel1}) nor (\ref{lel}) include
the region $R<2m$. Consider the analogy of the Schwarzschild solution.
Using isotropic coordinates, one can only access the region $R\geq 2m$
of the space-time. But then the coordinate transformation (\ref{req})
puts the line element into a form in which the extension into the region
$R<2m$ arises naturally. This involves skipping over a coordinate singularity
at $R=2m$, but nonetheless produces the line element for the region $R<2m$.
Here, the corresponding line element (\ref{lel}) {\em does not} apply to
the region $R<2m$. We must ask therefore if the solution does exist in this region, and if so, how the space-time metric and curvature extend through the singularity at $R=2m$.

Central to these questions is the issue of the gravitational strength of the singularity \cite{tipler}. There does not seem to be any way to extend through a sigularity which is classified as gravitationally strong (see below for what is meant by this); however we will show that the singularity at $R=2m$ is gravitationally weak, and thus the possibility of extending the space-time through this surface remains open.

We then ask what form the space-time region $R<2m$ might take.
We provide a partial and negative answer to this question by
showing that the hypotheses (i)-(iii) above may be cast in a
form which avoids any global assumptions (apart from that of
spherical symmetry), i.e. (ii) above is replaced with a local statement.
It then arises immediately that these equivalent conditions cannot be satisfied in the region $R<2m$. In this sense, McVittie's solution cannot be extended into the region $R<2m$.
This local characterization may be used to describe a point mass embedded in an arbitrary (open, flat or closed) RW universe.

We show that if the local characterization is applied to a {\em tachyon} fluid, then a solution does exist in, and only in, the region $R<2m$.

We conclude with some comments on the significance of these results.

\section{Gravitational strength of $R=2m$}

Two ways of characterizing the strength of a singularity are in
common usage: the {\em limiting focusing condition} of Krolak
\cite{krolak} and the {\em strong curvature condition} of Tipler
\cite{tipler}. We will use the latter. This is based on the idea
of classifying a singularity as strong if it crushes to zero
volume any extended body which falls into it. This is modeled in
the following way \cite{joshi}. Let $\gamma:[\tau_0,0)\rightarrow
M$ be a time-like geodesic which is incomplete at the parameter
(proper time along $\gamma$) value $\tau=0$. Let
$Z^{(\alpha)}_{a}, \alpha=1,2,3$ be linearly independent Jacobi
fields defined along $\gamma$ which (a) are orthogonal to $k^a$,
the unit tangent to $\gamma$ and (b) vanish at
$\tau=\tau_1\geq\tau_0$. These define a volume element along
$\gamma$, given by $V(\tau)=|\!|Z^{(1)}\wedge Z^{(2)}\wedge
Z^{(3)}|\!|$. Then $\gamma$ is said to terminate at a {\em strong
curvature singularity} if $\lim_{\tau\rightarrow 0}V(\tau)=0$ for
all possible choices of $Z^{(\alpha)}$ and $\tau_1$. A similar
characterization exists for null geodesics involving two
space-like Jacobi fields. Then a space-time singularity $p$ is
called {\em gravitationally strong} if every causal geodesic
terminating at $p$ terminates at a strong curvature singularity.
The singularity is called {\em gravitationally weak} if the volume
elements all have finite non-zero limits at the singularity.

We show that every radial causal geodesic terminating at $R=2m$ terminates at a weak curvature singularity, and so every point on $R=2m$ is gravitationally weak. We apply some general results on singularities in spherical symmetry \cite{ssss}.

Consider the time-like case. By linearity of the geodesic deviation equation,
which is satisfied by every Jacobi field, and the definition of $V(\tau)$
which involves the norm of a 3-form, $V(\tau)$ may be invariantly described
in terms of the norm $a$ of a radial space-like Jacobi field and the norms
$X^{(A)}, A=1,2$ of space-like Jacobi fields tangent to the metric 2-spheres
\cite{ssss}. The result is
\[ V(x)= aX^{(1)}X^{(2)}.\]
For a radial null geodesic, we obtain the same form for $V$ with $a$ set
equal to one.

It can be shown that the Jacobi conditions for $X^{(A)}$, for both the
null and time-like cases, lead to
\[ X^{(A)}(\tau)=R(\tau)\int_{\tau_1}^\tau \frac{d\nu}{R^2(\nu)},\]
where $X^{(A)}(\tau_1)=0$ (this is part of the definition of the Jacobi field).
Thus in the null case and for a non-central $(R>0)$ singularity, the weak curvature condition is always satisfied, and in the time-like case, $V\rightarrow 0$ if and only if $a(\tau)\rightarrow 0$. The Jacobi conditions for $a$ include
\begin{eqnarray}
{\ddot {a}}+ F(\tau)a=0,\label{jeq}
\end{eqnarray}
where an overdot indicates covariant differentiation along the geodesic in question and
\[ F(x)= 4\Psi_2 +2 \frac{E}{R^3} -\frac{Ric}{6}.\]
Here, $\Psi_2$ is the Newman-Penrose Weyl tensor term calculated on a null tetrad based on the principal null directions of the spherically symmetric space-time (in which sense it is invariant), $E$ is the Misner-Sharp gravitational energy and $Ric$ is the Ricci scalar. See \cite{ssss} for details.
The behaviour of $a$ in the limit $\tau\rightarrow 0$ is thus determined by the limiting behaviour of $F$ along the geodesic approaching $R=2m$. It only remains to determine this behaviour for an arbitrary time-like geodesic.

For $McV_4$, we have
\[ F(x) = -\frac{2m}{R^3}-\frac{\beta_t^2}{4}-\frac{\beta_{tt}}{2}\bigtriangleup^{-1/2},\]
where $\bigtriangleup=1-2m/R$.
Thus the factor determining the limiting behaviour of $a(\tau)$ as
$R\rightarrow 2m^+$ along the geodesic is $\bigtriangleup(\tau)$.
By definition, $\lim_{\tau\rightarrow 0}\bigtriangleup(\tau)=0$. What is important is the rate at which this approach to zero occurs. This is determined by the geodesic equations. It turns out that this information can be obtained quite easily.

The first integral of the radial time-like geodesic equations is
\[-(\bigtriangleup-\frac{1}{4}\beta_t^2R^2){\dot{t}}^2-\beta_tR\bigtriangleup^{-1/2}{\dot{t}}{\dot{R}}+\bigtriangleup^{-1}{\dot{R}}^2=-1.\]
Solving for ${\dot{t}}$, we can write
\[ -(\bigtriangleup-\frac{1}{4}\beta_t^2R^2){\dot {t}}=\frac{1}{2}\beta_tR\bigtriangleup^{-1/2}{\dot R}\pm({\dot R}^2+\bigtriangleup-\frac{1}{4}\beta_t^2R^2)^{1/2}.\]
This equation must always have a solution, and so the first integral {\em reveals} the constraint (substituting for $R$ in terms of $\bigtriangleup$)
\begin{eqnarray} 4m^2(1-\bigtriangleup)^{-4}{\dot \bigtriangleup}^2+
4\bigtriangleup-m^2\beta_t^2(1-\bigtriangleup)^{-2}&\geq&0.
\label{constraint}\end{eqnarray} We emphasize that this is not a
supplementary condition which we must impose on the functions
$R(\tau), t(\tau)$, but rather one which necessarily arises as a
consequence of the other geodesic equations. A similar situation
arises in Schwarzschild's space-time; there the constraint is
\[ m^2(1-\bigtriangleup)^{-4}{\dot{\bigtriangleup}}^2+\bigtriangleup \geq 0.\]
In the limit $\tau\rightarrow 0$, we know that $\bigtriangleup\rightarrow 0$, so that
\[ \lim_{\tau\rightarrow 0}{\dot\bigtriangleup}^2\geq \lim_{\tau\rightarrow 0}\frac{1}{4}\beta_t^2.\]
Along any particular geodesic, $\beta_t(\tau)$ remains finite
in the approach to the singularity.
This is a consequence of a result of paper II mentioned above, that
$R=2m$ cuts off the big bang (crunch) $\beta_t\rightarrow\infty$ from the rest of the space-time.
Also, $\beta_t\neq 0$, as we assume a non-static situation.
Thus either ${\dot\bigtriangleup}=O(1)$ as $\tau\rightarrow 0$, or ${\dot \bigtriangleup}$
becomes infinite in the limit, but with a finite integral.
Therefore we must have $\bigtriangleup=O(\tau^p)$ for some $p$,
$0<p\leq 1$. Then $\bigtriangleup^{-1/2}=O(\tau^{-p/2})$, and
the range of $p$ then indicates that
\[ \lim_{\tau\rightarrow 0} \tau^2F(\tau)=0.\]

This last result is sufficient to show that, quite generally, $a(\tau)$ is finite and non-zero in the limit $\tau\rightarrow 0$, yielding a weak curvature singularity along the geodesic in question.
See \cite{ssss} for details, and \cite{b+o} for the general theory.

Thus all Jacobi fields have finite norm as the singularity is approached along any radial causal geodesic, and so the singular surface $R=2m$ is classified as being gravitationally weak.

\section{A local characterization of the solution}
We have shown that the singular surface $R=2m$ is gravitationally weak. In paper II, we showed that it is space-like and trapped, and so using a triad of Jacobi fields at each point along the surface, it may be possible to use this as an initial data surface from which to evolve Einstein's equations into the region $R<2m$. We defer investigation of this possibility and instead attempt to solve Einstien's equations directly, subject to a set of conditions equivalent to (i)-(iii) above, but which apply throughout space-time. We find that it is impossible to do so.

Condition (i) above leads to the line element (see section 14.2 of \cite{ksmh})
\[ ds^2=-e^{\nu(r,t)}dt^2+e^{\mu(r,t)}(dr^2+r^2d\omega^2),\]
subject to
\begin{eqnarray}
e^\nu=\mu_t^2e^{g(t)},\qquad \mu^{\prime\prime}-\frac{1}{2}{\mu^\prime}^2-\frac{1}{r}\mu^\prime=F(r)e^{-\mu/2},
\label{feqs}
\end{eqnarray}
where the prime indicates partial differentiation with respect to $r$.
Spatial homogeneity (condition (iii)) is equivalent to $F(r)=6m/r^3$ for some constant $m$, which turns out to be the Schwarzschild mass of the vacuum limit. This is also equivalent to the renormalized Hawking mass \cite{hawking} of the space-time being equal to constant $m$. See paper I for details. This form of $F$ allows us to obtain the first integral of the second of (\ref{feqs}):
\begin{eqnarray}
rR^\prime = R(1-\frac{2m}{R}+ A(t)R^2)^{1/2},\label{mfeq}
\end{eqnarray}
where $R=re^{\mu/2}$ is the proper radius.
Then condition (ii) is satisfied if and only if the arbitrary function of integration $A(t)$ is equal to zero.

Leaving condition (ii) aside for the moment, we note that the function $A(t)$ appears in the energy density and the Misner-Sharp energy of the space-time: we have
\begin{eqnarray*} 8\pi\rho(t)&=&\frac{3}{4}e^{g(t)}-3A(t),\\
E&=&\frac{4}{3}\pi R^3\rho(t)+m.
\end{eqnarray*}
So neither of these terms isolates $A(t)$. However, it turns out that
\[ g^{ab}D_aRD_bR = 1-\frac{2m}{R}+A(t)R^2,\]
where $D_a=(g_{ab}+u_au_b)\nabla^b$ is the spatial gradient orthogonal to the fluid flow, which has unit time-like tangent $u^a=e^{-\nu/2}\delta^a_t$. This quantity, and the terms $m$ and $R$ on the right hand side are covariantly defined, and so we have a covariant definition of $A(t)$. So we can replace the asymptotic condition (ii) with the {\em local} condition
\begin{eqnarray}
{\rm{(ii)}}^\prime \qquad\qquad
g^{ab}D_aRD_bR = 1-\frac{2m}{R}.\label{new2}
\end{eqnarray}
Since the spatial gradient of any quantity is space-like, we see that we must have $R\geq 2m$. Thus we have demonstrated the following result:
\newline
{\bf Proposition:}
{\em There is no equivalent of McVittie's solution in the region $R<2m$.}

The local characterization above applies to the case $k=0$, but
may be generalized to $k$ positive and negative. In the case
$k=-1$, we showed in paper I that subject to conditions (i) and
(iii) above, the appropriate value for the function of integration
$A(t)$ generated by an asymptotic condition is $e^{-\beta(t)}$,
where $S(t)=e^{\beta/2}$ is the scale factor of a RW universe with
negatively curved spatial sections. This leads to
\[ g^{ab}D_aRD_bR = 1-\frac{2m}{R}+e^{-\beta}R^2.\]
On the other hand, in an arbitrary RW universe we have
\[g^{ab}D_aRD_bR = 1-ke^{-\beta}R^2.\]
Thus we can change ${\rm{(ii)}}^\prime$ so that it applies to the embedding of a point mass in any RW universe:
\begin{eqnarray}
{\rm{(ii)}}^{\prime\prime} \qquad\qquad
g^{ab}D_aRD_bR = 1-\frac{2m}{R}-ke^{-\beta}R^2.\label{2new2}
\end{eqnarray}
Then conditions (i), (ii)$^{\prime\prime}$ and (iii) describe the space-time representing a point mass $m$ embedded in a RW universe with curvature index $k$ and scale factor $e^{\beta/2}$.

Unfortunately, the solution in the cases $k\neq 0$ cannot be obtained in
closed form, as (\ref{2new2}) leads to an elliptic integral. Nevertheless,
we were able to demonstrate in paper I several of the properties of the
$k=-1$ solution. Among these was that the vacuum limit yields, as one would
expect, the Schwarzschild solution. A serious doubt is cast on the point
mass interpretation for the $k=+1$ solution by the fact that this limit
does not exist in this case. The energy density of the space-time is given by
\[ 8\pi\rho(t)=\frac{3}{4}\beta_t^2+3ke^{-\beta},\]
which for $k=+1$ cannot be set equal to zero,
as it can for $k=0$ and $k=-1$, in which cases the
pressure also automatically vanishes, yielding the
spherical vacuum with non-zero Weyl tensor, i.e.
Schwarzschild space-time. The inability to obtain the
Schwarzschild limit is equivalent to the fact that flat
space-time is not a special case of a $k=+1$ RW universe.
Indeed this fact indicates that one cannot embed the Schwarzschild
solution into a $k=+1$ RW
 universe in any physical way.
In such a situation, one must be able to identify in a gauge invariant
manner the Schwarzschild part of the solution, which would
involve looking for a well defined vacuum limit. If one could
produce the Schwarzschild solution in this way, then by setting
$m=0$ one would also be able to produce flat space-time, which, as we have just seen, is impossible.

\section{A tachyon fluid in $R<2m$}
Two facts suggest that a tachyon fluid might be an appropriate matter distribution for the region $R<2m$.

Firstly, what hinders the existence of the solution in $R<2m$ is the fact that one of the conditions we are trying to satisfy is $g^{ab}D_aRD_bR = 1-\frac{2m}{R}$, where $D_a$ is the spatial gradient orthogonal to the fluid flow. Obviously if this were a temporal gradient orthogonal to a unit space-like vector, the situation would be reversed; the equation could only be satisfied in the region $R\leq 2m$.

Secondly, in the Schwarzschild solution, the preferred vector field $\partial/\partial t$ is time-like in the region $R>2m$, but space-like in $R<2m$. This suggests that something similar may occur in $McV_4$; the preferred vector field - the eigen-vector of the energy-momentum tensor whose eigen-value has multiplicity one - switches over from being time-like to being space-like. We investigate this possibility here, and show that the space-time which arises uniquely is remarkably similar to $McV_4$.

Thus we consider a space-time described by conditions (i)-(iii) below.

\newcounter{tachyon}
\begin{list}
{(\roman{tachyon})}{\usecounter{tachyon}}
\item
{\em Space-time $(M,g)$ is spherically symmetric and
filled with a shear-free tachyon fluid.}
\end{list}

The condition on the energy-momentum tensor is thus
\[ T_{ab}=(\rho-p)n_an_b+pg_{ab},\]
where $g_{ab}n^an^b=1$, and we refer to $\rho$ and $p$ as the tachyon energy density and pressure. The shear of a congruence of space-like curves (the tachyon flow lines) is defined in exactly the same way as that of a congruence of time-like curves.
We can use co-moving coordinates and write the line-element as
\[ds^2=e^\nu dt^2+e^\mu(-dr^2+r^2d\omega^2),\]
where $\nu=\nu(r,t)$, $\mu=\mu(r,t)$ and the shear-free condition has been incorporated. The unit tangent to the tachyon flow is $n^a=e^{-\nu/2}\delta^a_t$.
The vanishing of non-diagonal terms in $T_{ab}$ and the tachyon pressure isotropy lead to
\begin{eqnarray} e^\nu&=&\mu_t^2e^{g(t)},\label{teq1}\\
\mu^{\prime\prime}-\frac{1}{2}{\mu^\prime}^2-\frac{1}{r}\mu^\prime-\frac{4}{r^2}&=&F(r)e^{-\mu/2}.\label{teq2}
\end{eqnarray}

\begin{list}
{(\roman{tachyon})}{\usecounter{tachyon}}
\setcounter{tachyon}{1}
\item
{\em The tachyon energy density is homogeneous; $D_a\rho:=(g_{ab}-n_an_b)\nabla^b\rho=0$.}
\end{list}

This is equivalent to $\rho=\rho(t)$, and as in the fluid case, leads to the equation $F=-c/r^3$ for some constant $c$ which will be identified later. Then (\ref{teq2}) may be integrated once to obtain
\[g^{ab}D_aRD_bR=1-\frac{2}{3}cR^{-1}-B(t)R^2.\]
The term $B(t)$ appears in the tachyon energy density;
\[8\pi\rho=\frac{3}{4}e^{-g(t)}-3B(t).\]
By analogy with the fluid case, we take $B=0$ to be the third condition. This condition guarantees that the $c=0$ background ($c$ will be identified as being proportional to the Schwarzschild mass) is an isotropic tachyon fluid with Lorentzian sections $t=$constant of zero curvature. The condition could be modified to include constant non-zero curvature, but we wish to focus on the case analogous to $k=0$.

\begin{list}
{(\roman{tachyon})}{\usecounter{tachyon}}
\setcounter{tachyon}{2}
\item
{\em The gradient of the radius function orthogonal to the tachyon flow obeys $g^{ab}D_aRD_bR=1-\frac{2}{3}cR^{-1}$.}
\end{list}

This last equation may be integrated to obtain
\[e^{\mu/2}=\frac{c}{3r}(1+\sin{(a(t)+\ln{r})}),\]
where $a(t)$ is arbitrary. We can then calculate
\[e^\nu=4a_t^2(t)e^{g(t)}(\frac{2c}{3R}-1).\]
Rescaling $t$ allows us to set $4a_t^2(t)e^{g(t)}=1$, and we define $\alpha(t)=2a(t)$.
Carrying out the transformation $R(r,t)=re^{\mu/2}$ puts the line element into the form
\[ ds^2=(-1+\frac{2c}{3R}-\frac{1}{4}\alpha_t^2R^2)dt^2+\alpha_tR(\frac{2c}{3R}-1)^{-1/2}dRdt
-(\frac{2c}{3R}-1)^{-1}dR^2+R^2d\omega^2.\]
The vacuum limit must give the Schwarzschild solution. We can show that this limit is $\alpha=$constant, which gives the line element
\[ ds^2=(-1+\frac{2c}{3R})dt^2
-(\frac{2c}{3R}-1)^{-1}dR^2+R^2d\omega^2,\]
and so we identify $c=3m$, where $m$ is the Schwarzschild mass parameter. Then the line element reads
\begin{eqnarray}
 ds^2=(-1+\frac{2m}{R}-\frac{1}{4}\alpha_t^2R^2)dt^2+\alpha_tR(\frac{2m}{R}-1)^{-1/2}dRdt
-(\frac{2m}{R}-1)^{-1}dR^2+R^2d\omega^2.
\label{tachsol}
\end{eqnarray}
The tachyon energy density and pressure are given by
\[ 8\pi\rho=\frac{3}{4}\alpha_t^2,\qquad 8\pi p=\frac{3}{4}\alpha_t^2+\alpha_{tt}(\frac{2m}{R}-1)^{-1/2}.\]

The solution only exists in the region $R<2m$. The limit $m=0$
yields the unique isotropic tachyon fluid solution with Lorentzian
sections orthogonal to the tachyon flow of zero curvature. The
vacuum limit $\alpha=$ constant yields the Schwarzschild solution.
Setting $\rho+p=0$ in $McV_4$, i.e. considering the special case
of an Einstein space, we obtain the Schwarzschild-de Sitter
universe. It is interesting to note that for the tachyon fluid, we
obtain the Schwarzschild-anti de Sitter universe. For
$T_{ab}\propto g_{ab}$ is equivalent to $\rho-p=0$ for the tachyon
fluid, corresponding here to $\alpha_t=A=$constant. The metric
tensor may then be diagonalized using a transformation $T=t+f(R)$,
with $f$ chosen to remove off-diagonal terms. The resulting line
element is
\[ ds^2=-(1-\frac{2m}{R}+\frac{A^2}{4}R^2)dT^2+(1-\frac{2m}{R}+\frac{A^2}{4}R^2)^{-1}dR^2 +R^2d\omega^2.\]
This is the Schwarzschild-anti de Sitter line element with negative cosmological constant $\Lambda=-3A^2/4$.

\section{Discussion}
We have highlighted here a curious aspect of McVittie's solution, namely that it displays an unstable aspect of the Schwarzschild event horizon and the black hole interior. Physical perturbations of the event horizon are governed by Price's theorem \cite{mtw} and indicate its stability. However these perturbations do not extend to the situation considered here, where the black hole is immersed in an isotropic perfect fluid. The result is that the event horizon becomes singular, albeit weakly singular. An analogous situation obtains in the interior; the black hole interior cannot be embedded into an isotropic perfect fluid in the manner described above. It is therefore important to determine the following. Is this behaviour generic in any way ? That is, is McVittie's solution just one of a general class of solutions representing the physical embedding of the Schwarzschild space-time into a perfect fluid in which the event horizon becomes singular, or is this behaviour specific to McVittie's solution? Furthermore, one should ask if such models have any basis in reality; could such space-times be used to model matter accreting onto a black hole, for instance? With regard to the first question, removing the simplifying assumption of shear-free fluid flow, as would be necessary to discuss the more general situation, greatly increases the complexity of the field equations (see paper II for related comments). However the conditions described in section one above offer an indication of what boundary conditions should be used in solving Einstein's equations to address this question and these could lead to some simplifications.

We have given an unphysical solution which applies to the embedding of the
black hole interior into an isotropic medium. To obtain a physically (more)
realistic solution, it would be helpful to find a geometric extension of the
metric (\ref{lel}) through the weak
space-like singular hypersurface $R=2m$.
The fact that the singularity is weak indicates that it may be possible to
define, in some invariant or canonical way, a metric at $R=2m$, using the
full set of finite Jacobi fields at the surface. If it were possible to do
the same for the first fundamental form at $R=2m$, then it may be possible
to set up a well posed initial value problem for the metric inside $R<2m$.
These ideas are very loosely stated, but the simple form of the singularity
in this case may allow some progress to be made in this direction. We note
that a good deal of progress has been made on a somewhat similar program
which deals with space-times containing a different class of singularities
which have a manageable degree of pathology, namely isotropic
(or conformally compactifiable) singularities \cite{tod}.
For certain types of energy-momentum tensor, the conformal
Einstein equations for such space-times have been shown to
possess a well-posed Cauchy problem with initial data consisting
of the conformal 3-metric at the singularity \cite{anguige+tod}.

We conclude by giving the latest version of the interpretation of
McVittie's solution. The line element (\ref{lel1}) with $u>m/2$
gives the solution for the exterior region $R>2m$ of the Schwarzschild
black hole embedded into a flat isotropic universe with scale-factor
$e^{\beta/2}$; the interior region cannot be embedded in this way.

\end{document}